\documentclass[%
 reprint,
 superscriptaddress,
 amsmath,amssymb,
 aps,
]{revtex4-1}

\usepackage{graphicx}
\usepackage{dcolumn}
\usepackage{bm}
\usepackage{xcolor}

\usepackage{subfigure}

\begin{document}


\title{Simulation of dark scalar particle sensitivity
in $\eta$ rare decay channels at HIAF}    

\author{Yang Liu}
\affiliation{Institute of Modern Physics, Chinese Academy of Sciences, Lanzhou 730000, China}
\affiliation{School of Nuclear Science and Technology, University of Chinese Academy of Sciences, Beijing 100049, China}

\author{Rong Wang}
\email{Corresponding author: rwang@impcas.ac.cn}
\affiliation{Institute of Modern Physics, Chinese Academy of Sciences, Lanzhou 730000, China}
\affiliation{School of Nuclear Science and Technology, University of Chinese Academy of Sciences, Beijing 100049, China}

\author{Zaiba Mushtaq}
\affiliation{Institute of Modern Physics, Chinese Academy of Sciences, Lanzhou 730000, China}
\affiliation{School of Nuclear Science and Technology, University of Chinese Academy of Sciences, Beijing 100049, China}

\author{Ye Tian}
\affiliation{Institute of Modern Physics, Chinese Academy of Sciences, Lanzhou 730000, China}
\affiliation{School of Nuclear Science and Technology, University of Chinese Academy of Sciences, Beijing 100049, China}

\author{Xionghong He}
\affiliation{Institute of Modern Physics, Chinese Academy of Sciences, Lanzhou 730000, China}
\affiliation{School of Nuclear Science and Technology, University of Chinese Academy of Sciences, Beijing 100049, China}

\author{Hao Qiu}
\affiliation{Institute of Modern Physics, Chinese Academy of Sciences, Lanzhou 730000, China}
\affiliation{School of Nuclear Science and Technology, University of Chinese Academy of Sciences, Beijing 100049, China}

\author{Xurong Chen}
\affiliation{Institute of Modern Physics, Chinese Academy of Sciences, Lanzhou 730000, China}
\affiliation{School of Nuclear Science and Technology, University of Chinese Academy of Sciences, Beijing 100049, China}


\date{\today}

\begin{abstract}
Searching dark portal particle is a hot topic in particle physics frontier.
We present a simulation study of an experiment targeted for searching
the scalar portal particle at Huizhou $\eta$ factory.
The HIAF high-intensity proton beam and a high event-rate spectrometer are
suggested for the experiment aimed for the discovery of new physics.
Under the conservative estimation, $5.9\times 10^{11}$ $\eta$ events
could be produced in one month running of the experiment.
The hadronic production of $\eta$ meson ($p + ^7\text{Li} \rightarrow \eta X$) is
simulated at beam energy of 1.8 GeV using GiBUU event generator.
We tend to search for the light dark scalar particle in the rare decay channels
$\eta \rightarrow S \pi^0 \rightarrow \pi^+ \pi^- \pi^0$
and $\eta \rightarrow S \pi^0 \rightarrow e^+ e^- \pi^0$.
The detection efficiencies of the channels and
the spectrometer resolutions are studied in the simulation.
We also present the projected upper limits of the decay
branching ratios of the dark scalar particle
and the projected sensitivities to the model parameters.
\end{abstract}

\maketitle


\section{Introduction}
\label{sec:intro}

Over the past few decades, identifying the Dark Matter (DM) and Dark Energy
has been one of the most extensively researched topics in particle physics and astrophysics
\cite{Young:2016ala,Arbey:2021gdg,Oks:2021hef,Bertone:2018krk,Aramaki:2015pii}.
The Standard Model (SM) of particle physics is precisely tested,
however it fails in explaining the mechanism of dark matter production
and the possible interactions of DM with the ordinary visible matter.
Pursuing the new physics, the theoretical physicists have been continuously
proposing new ideas of extending the SM
\cite{Lanfranchi:2020crw,Gan:2020aco,Holdom:1985ag,Galison:1983pa,Fayet:1990wx,
OConnell:2006rsp,Krnjaic:2015mbs,Batell:2018fqo,Batell:2017kty,Georgi:1986df,Gorbunov:2007ak},
and at the same time many experiments have been conducted
in the hope of unraveling the mysteries of DM
\cite{Zhao:2020ezy,Liu:2017drf,Cebrian:2022brv,ADMX:2018gho,CDEX:2023wfz,
Hochberg:2022apz,Georgescu:2022npr}.
For an instance, there have been some high-energy experiments aimed
to probe the interactions between DM and visible matter
at the Large Hadron Collider, and possibly
to measure the properties of the DM particles once detected
\cite{Buchmueller:2017qhf,Felcini:2018osp}.

Among the theories proposed in the past, Weakly Interacting Massive Particle
(WIMP) was considered one of the promising candidates for DM
\cite{Jungman:1995df,Bertone:2004pz,Feng:2010gw}.
Typically, it has a very large mass scale ($\sim$ TeV) and only interacts
with visible matter via the weak interaction.
However, the parameter space of WIMP is almost experimentally excluded
\cite{ParticleDataGroup:2024cfk,Liu:2017drf,Roszkowski:2017nbc,CDMS:2000kdm}.
Theoretically, the light WIMP particles (MeV $\sim$ GeV),
would be produced very abundantly in the early universe,
unless their annihilation rates are accelerated by the accompanying production
of neutral mediator particles \cite{Lanfranchi:2020crw,Pospelov:2007mp,Pospelov:2008jd}.
As the light DM particle currently has few experimental constraints,
it has become an important direction for future experimental searches.
More and more collaborations worldwide are turning
to search for the DM particles in low-mass region
\cite{Lanfranchi:2020crw,Gan:2020aco,Baltzell:2022rpd,REDTOP:2022slw}.

The interaction mediators in the dark sector have small masses (MeV $\sim$ GeV)
compared to WIMP, hence they are the relatively long-lived,
electrically neutral vector or scalar particles \cite{Lanfranchi:2020crw,Pospelov:2007mp,Pospelov:2008jd}.
Recently, there are some experimental data indicating
the presence of new physics in the low-energy region around GeV scale
\cite{Krasznahorkay:2015iga,Muong-2:2023cdq,Muong-2:2021ojo,Miller:2007kk},
which can be explained by involving the interactions
with some invisible particles via the so-called ``portal'' particles
\cite{Feng:2016jff,Feng:2016ysn,NA64:2021acr,Bodas:2021fsy,Nomura:2020kcw}.
Unlike the traditional WIMP, the portal particle is not
required to has a large mass or to contribute significantly to DM.
The portal particles act just as interaction mediators
between the dark sector and the SM sector.
They are probably the new gauge particles coupling
feebly with the visible matter \cite{Lanfranchi:2020crw,FASER:2018eoc}.
According to their quantum numbers, they are usually classified
into the vector portal particles, the scalar portal particles,
the heavy lepton portal particles, and the axion-like portal particles.

As a quasi Goldstone particle, $\eta$ is of all zero quantum numbers ($I^{G}J^{PC}=0^+0^{-+}$).
The $\eta$ meson has relatively small decay width,
since many strong and electromagnetic decays are forbidden at the tree level,
due to the conservations of P, C, G parities and angular momentum.
Therefore, the $\eta$ rare decay channels involving the dark portal particles
have relatively large decay widths.
The $\eta$ meson was also recognized long ago as a testing ground
for discrete symmetry violation \cite{Prentki:1965tt}.
As much interesting physics is involved in the rare decay channels of $\eta$ meson,
there are already many experiments aimed at the precise measurements of $\eta$ decays,
such as BESIII \cite{BESIII:2012nen,BESIII:2023edk,BESIII:2015fid,BESIII:2012tzz,BESIII:2014bgm,BESIII:2016tdb},
KLOE/KLOE-II \cite{Krzemien:2019ktq,KLOE-2:2020ydi,KLOE-2:2016zfv,KLOE:2008arm},
and JLab Eta Factory experiments \cite{Gan:2020aco}.
Recently, the REDTOP experiment is proposed to probe new physics
via the rare $\eta$ and ${\eta^{\prime}}$ decays,
with much more statistics of the $\eta$ yield \cite{REDTOP:2022slw}.
It is suggested that the $\eta$ meson decay is highly
suitable for studying many conjectured dark portal particles
which connect the SM sector to the hidden sector.

To explore the new physics via $\eta$ rare decays,
the statistic of $\eta$ mesons needs to be extremely high.
Since the hadronic production reaction of $\eta$ meson
is of significantly large cross section,
the high-intensity proton beam is an ideal tool
to yield the unprecedented number of $\eta$ samples.
Recently, a plan of building a super $\eta$ factory at Huizhou is suggested \cite{Chen:2024wad}. 
We will provide much more information on simulation of dark scalar particle 
at the suggested $\eta$ factory at Huizhou than that in the Ref. \cite{Chen:2024wad}.
The High Intensity heavy-ion Accelerator Facility (HIAF)
currently under construction at Huizhou provides a great opportunity
of building such a super $\eta$ factory,
as HIAF can offer the ion beam of the strongest
pulse intensity in its energy region.
To see the physics impact on the dark scalar particle
at the suggested Huizhou $\eta$ factory,
we have performed some simulations.
In this work, we show the details of the simulation
and the projected sensitivity to the dark scalar particle
for a prior experiment of one month running at Huizhou $\eta$ factory.

The organization of the paper is as follows.
A brief review of the theoretical models of dark
scalar portal particles is given in Sec. \ref{sec:theory}.
The conceptual design of the spectrometer
for the super $\eta$ factory is discussed in Sec. \ref{sec:spectrometer}.
The simulation framework for this study is introduced Sec. \ref{sec:simulation}.
The simulation results and some discussions are given in Sec. \ref{sec:result}.
At the end, a concise summary is provided in Sec. \ref{sec:summary}.

\section{Theoretical models of dark scalar portal}
\label{sec:theory}

In the scalar portal models, the dark sector couples to
the SM sector via the interaction with the Higgs boson
or an extension of the latter \cite{Burgess:2000yq,Piazza:2010ye,Kouvaris:2014uoa,
Pospelov:2007mp,Krnjaic:2015mbs}.
A new scalar particle is conjectured in these models.
It is often called the dark scalar particle
as it also couples to the hidden dark sector.
In this simulation, we focus on testing the minimal scalar model
and the hadrophilic scalar model, which allow
the rare decay channels $\eta\rightarrow S\pi^0\rightarrow e^+e^-\pi^0$
and $\eta\rightarrow S\pi^0\rightarrow\pi^+\pi^-\pi^0$ respectively.

\subsection{Minimal Scalar Model}

The simplest extension of the scalar sector of SM \cite{OConnell:2006rsp,
Burgess:2000yq,Patt:2006fw,Batell:2009di,Feng:2017vli} is
to add a single real scalar $S$ that is a gauge singlet.
As the minimum extension to the scalar field of SM,
the model is characterized by the inclusion of an additional singlet field,
and the presence of two types of couplings, namely $\mu$ and $\lambda$ \cite{OConnell:2006rsp}.
At the low energies, the involved dark scalar particle decays into
the electron-positron pair in these models \cite{Feng:2017vli,Bezrukov:2013fca}.
Therefore, one could find the signal of a light dark scalar particle
in the $\eta$ decay channel $\eta\rightarrow S\pi^0\rightarrow e^+e^-\pi^0$.
The SM decay of $\eta\rightarrow e^+e^-\pi^0$ is usually described with
a two-photon intermediate state to conserve C parity.
The branching ratio of the SM decay $\eta\rightarrow e^+e^-\pi^0$
is estimated to be on the order of $10^{-9}$, contributing a small SM background.
The relevant parameter of dark scalar particle in $\eta$ decay is the mixing angle $sin(\theta)$,
which describes the mixing effect of Higgs boson and dark scalar particle.
At low energies, the Higgs field can be described with $H=(v+h)/\sqrt{2}$,
where $v$ is the Electric-Weak vacuum expectation value
and $h$ is the field corresponding to the physical Higgs boson.
The nonzero $\mu$ in the dark scalar portal $\mu S HH$ leads to
the small mixing between Higgs boson and dark scalar particle,
which is written as $\theta = \mu v/(m_{\rm H}^2-m_{\rm S}^2)$
in small-mixing limit \cite{Lanfranchi:2020crw}.
Following Tulin's parametrization \cite{Gan:2020aco,REDTOP:2022slw},
the mixing angle is connected with
the branching ratio of $\eta$ decay via the following equation:
\begin{equation}
\begin{split}
{\rm Br}(\eta\rightarrow\pi^0 S) \simeq 1.8\times10^{-6}
\lambda^{1/2}\left(1,\frac{m^2_S}{m^2_{\eta}},\frac{m^2_{\pi^0}}{m^2_{\eta}}\right)
\sin^2{\theta},
\end{split}
\label{eq:BrAndTheta}
\end{equation}
where the mixing angle $\theta$ is an unknown parameter,
and $\lambda$ is a function related to the kinematics
($\lambda(a,b,c) = a^2 + b^2 + c^2 - 2ab - 2ac - 2bc$).
Note that Eq. (\ref{eq:BrAndTheta}) is based on a numerical
and rough evaluation with the uncertainty at order of 20\%.

\subsection{Hadrophilic Scalar Model}

Recently, a hardrophilic (or leptophobic) scalar model has been brought up.
Nonetheless, it encounter a set of challenges,
including a new flavor-changing neutral currents (FCNC) and
a naturalness problem related to the light scalar mass \cite{Batell:2017kty,Batell:2018fqo}.
In general, to meet the constraints of FCNC,
the hardrophilic scalar interaction must be ``flavor-specific''.
Some extensive analyses have been done on these issues \cite{Batell:2017kty,Batell:2018fqo},
in which it was found that the couplings to only specific quark mass eigenstates
can satisfy the existing FCNC constraints even for relatively large couplings
in the natural parameter space region.
This assumption provides a promising way for searching the hardrophilic scalar particle,
and the only coupling to the first-generation quarks is considered in this work.
The corresponding Lagrangian is written as,
\begin{equation}
\begin{split}
\mathcal{L} \supset \frac{1}{2}(\partial_{\mu}S)^2 - \frac{1}{2}m^{2}_{\rm S}S^{2} - g_u S \bar{u}u,
\end{split}
\label{eq:LagrangianHadrophilicModel}
\end{equation}
where $m_{\rm h}$ is the scalar mass,
and $g_u$ is the effective coupling parameter to the up quark.
Under the hadrophilic scalar model, the branching ratio of dark scalar particle
in $\eta$ rare decay is expressed as \cite{Batell:2018fqo,Gan:2020aco,REDTOP:2022slw},
\begin{equation}
\begin{split}
{\rm Br}(\eta\rightarrow\pi^0 S) = \frac{c^2_{S\pi^0\eta} g^2_u B^2} {16\pi m_{\eta} \Gamma_{\eta}}
\lambda^{1/2}\left(1, \frac{m^2_S}{m^2_{\eta}}, \frac{m^2_{\pi^0}}{m^2_{\eta}}\right),
\end{split}
\label{eq:BrAndgu}
\end{equation}
where $g_u$ is an unknown coupling parameter,
$B\cong m^2_{\pi}/(m_u+m_d) \approx 2.6$ GeV,
$c_{S\pi^0\eta} = \frac{1}{\sqrt{3}}\cos{\theta}-\sqrt{\frac{2}{3}}\sin{\theta}$
is a parameterized coefficient used to describe the effect of $\eta -\eta^{\prime}$ mixing,
and $\lambda$ is a function related to the kinematics.
Due to $SU(3)$ breaking, the physical states of $\eta$ and $\eta^{\prime}$ mesons are
the mixed states of the singlet and octet states.
In studying the meson decay at the quark level,
the $\eta -\eta^{\prime}$ mixing angle is considered.
At low energies, the interactions of dark scalar particle with pseudo Nambu-Goldstone
boson is associated with chiral symmetry breaking in terms of the known meson masses,
resulting in a dimensional parameter $B\simeq m^2_{\pi}/(m_u+m_d)$ \cite{Batell:2018fqo}.

\section{A compact silicon-pixel-based spectrometer}
\label{sec:spectrometer}

\begin{figure*}[htbp]
\centering  
\includegraphics[width=0.65\textwidth]{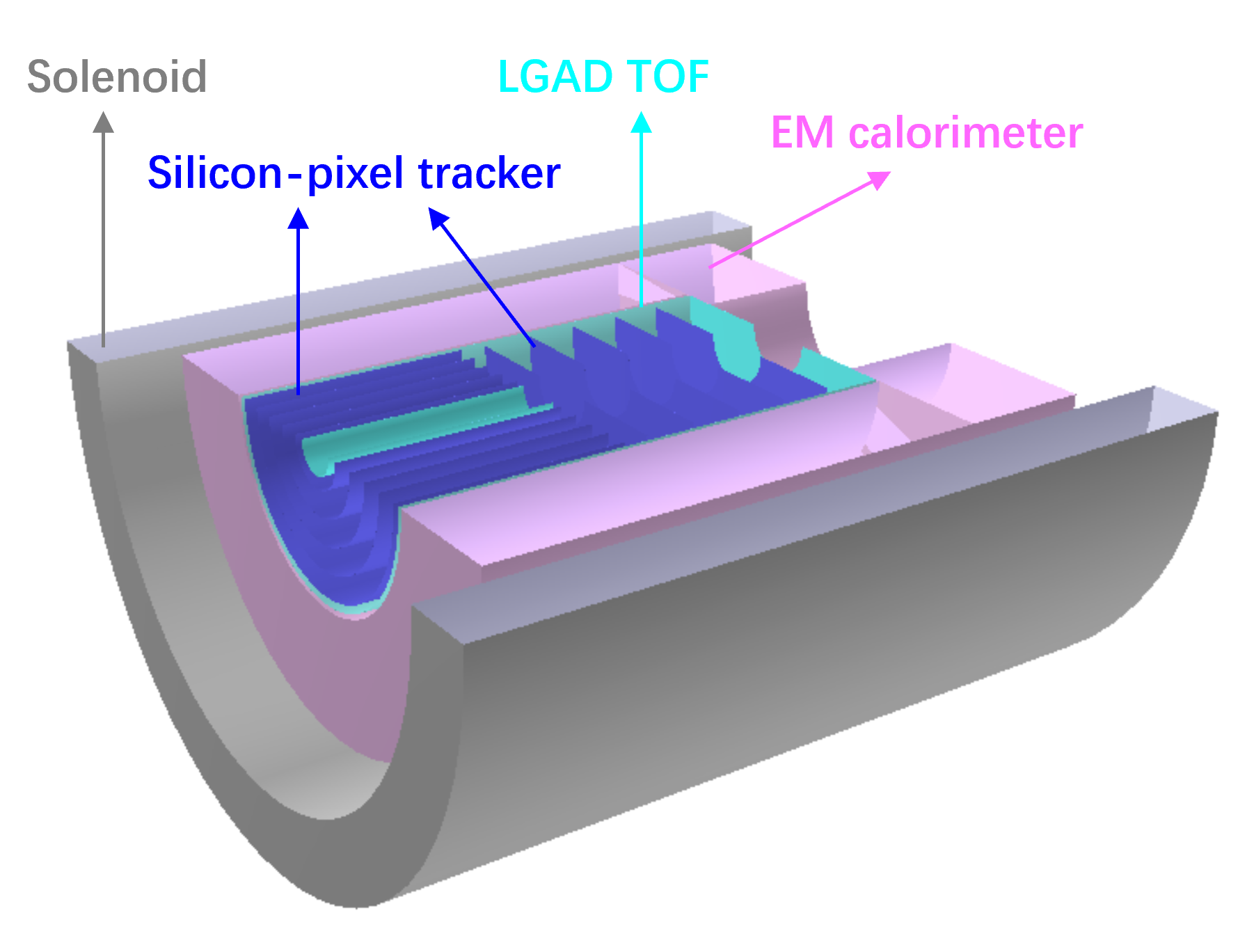}
\caption{The conceptual design of a compact spectrometer
for the $\eta$ rare decay experiment.
The grey, magenta, cyan, and blue modules show
the solenoid, EM calorimeter, TOF detector, and silicon-pixel tracker, respectively.
}
\label{fig:spectrometer}
\end{figure*}

The conceptual design of the spectrometer for the super $\eta$
factory experiment is shown in Fig. \ref{fig:spectrometer},
with the main parts indicated in the figure.
The tracking system is fully based on the silicon pixel detectors
of small position resolution close to 10 $\mu$m,
with forward parallel plate modules separated by 10 cm
and central barrel modules of 5 cm gaps.
The event-rate capacity is guaranteed with the dual
readout technique of both the arrival time and the deposited energy of the pixel.
The time-of-flight (TOF) detector is made of low gain avalanche detector (LGAD)
of low material budget, acting as the main particle identification detector
for low-energy particles. The outer layer TOF detector has the length
of 100 cm and the radius of 30 cm.
Outside the TOF detector, there is the electro-magnetic calorimeter (EMC)
made of radiation-hard lead glass, which is used for high-energy photon detection.
The fast time response of the Cerenkov light in lead glass enables
the high event rate of EMC. Moreover, the abundant
neutron background is significantly suppressed,
owing to the very few Cerenkov lights in the hadronic shower.
Our geant4 simulation implies that the low-energy neutron background
can be neglected in the lead-glass EMC.
All the main detectors mentioned above are inside a superconducting solenoid,
so as to measure the momentum of the charged particle.

\begin{figure*}[htbp]
\centering  
\includegraphics[width=0.65\textwidth]{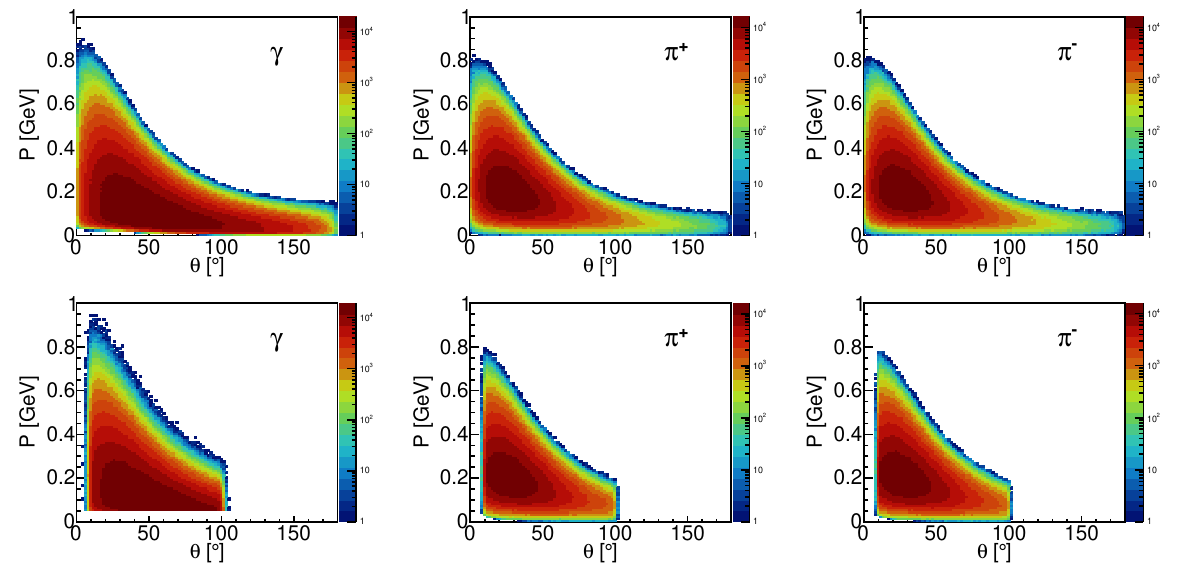}
\caption{The momentum versus angle distributions of the final-state particles
from the decay channel $\eta\rightarrow \pi^{+} \pi^{-} \pi^{0} (\gamma\gamma)$. }
\label{fig:momentum_angle_distributions_3pi_channel}
\end{figure*}

\begin{figure*}[htbp]
\centering  
\includegraphics[width=0.65\textwidth]{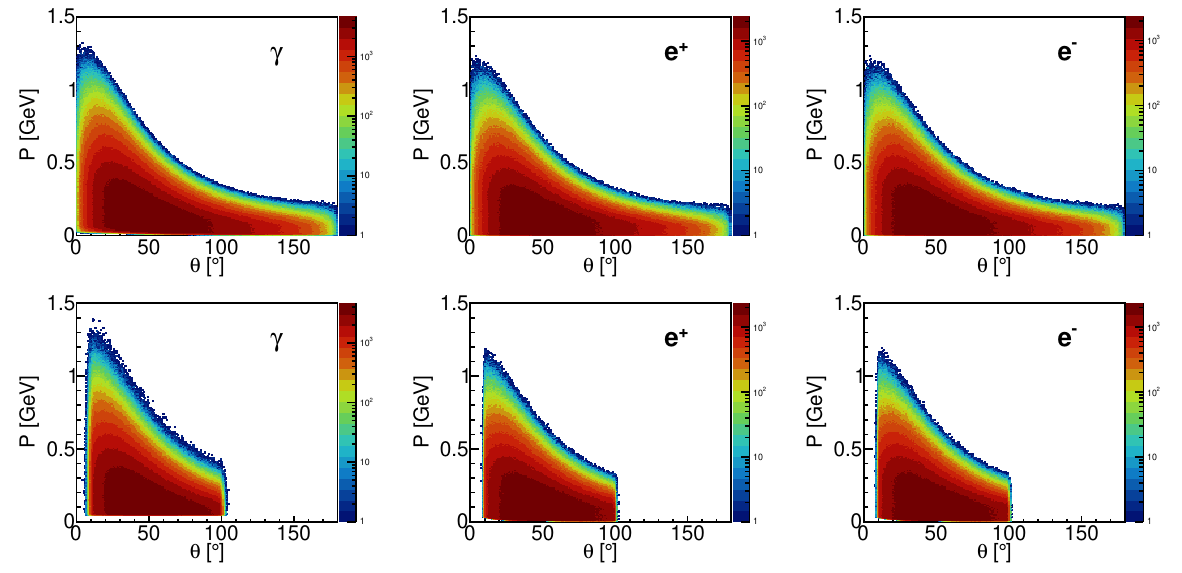}
\caption{The momentum versus angle distributions of the final-state particles
from the decay channel $\eta\rightarrow e^{+} e^{-} \pi^{0} (\gamma\gamma)$. }
\label{fig:momentum_angle_distributions_EEPi_channel}
\end{figure*}

Thanks to the small position resolution and high event-rate capacity of
the silicon pixel detector, the spectrometer is made compact,
with the solenoid inner radius of about 60 cm.
The multi-layer thin foil targets of light nuclei are placed
around the entrance of the spectrometer,
to achieve a large acceptance of the forward particles
in the fixed-target experiment.
The momentum and angular distributions of the final-state particles
of the aimed $\eta$ decay channels are shown in Fig. \ref{fig:momentum_angle_distributions_3pi_channel}
and \ref{fig:momentum_angle_distributions_EEPi_channel}.
The momentum and angular distributions of the reconstructed final-state particles
from the detector simulation are also shown in the figures.
We find that the majority part of the final-state particles
can be effectively measured with the currently designed spectrometer.

\section{Simulation Framework}
\label{sec:simulation}

For the $\eta$ meson production in proton-nucleus collision,
we adopt the Giessen Boltzmann-Uehling-Uhlenbeck (GiBUU) event generator,
which is based on the Boltzmann equation and the Uehling-Uhlenbeck equation
\cite{Buss:2011mx,Gaitanos:2007mm,Weil:2012ji},
for studying particle interactions and transport processes in a nuclear environment.
The interactions, collisions, and scattering processes between the nucleons
are taken into account with the Monte-Carlo method for the numerical solutions.
The GiBUU generator is applicable for describing various nuclear physics phenomena
in the energy range from 100 MeV to 100 GeV,
such as the heavy-ion collision, the nuclear reaction, and the nuclear structure study.
Following the $\eta$ meson production by GiBUU event generator,
some decay chains of $\eta$ meson are programmed by us,
in order to ensure a more realistic simulation and to study the
efficiency and resolution of the channel of interests.

For the spectrometer simulation, we construct the
ChnsRoot package that is based on the FairRoot framework \cite{Al-Turany:2012zfk}.
The FairRoot framework encompasses the core services
for detector simulation and offline analysis,
which enables the users to construct the experimental setups quickly and conveniently.
In ChnsRoot, we implement the fast simulation based on the geant4 simulation results
on the energy resolutions and efficiencies of the detectors.
With the ChnsRoot package, we can efficiently and reliably study
the acceptances, efficiencies and resolutions of the detectors.

The angular acceptances of both charged and neutral particles are designed
to be from $10^{\circ}$ to $100^{\circ}$, for the conceptual design
of the spectrometer. In Figs. \ref{fig:momentum_angle_distributions_3pi_channel}
and \ref{fig:momentum_angle_distributions_EEPi_channel},
we show the reconstructed kinematics from the ChnsRoot simulations
of pion, electron and photon in the studied $\eta$ decay channels.
The measured minimum momentum of charged particle is limited
by the inner radius of the silicon pixel tracker.
The hit threshold of EMC for neutral particle is set at the value
induced by a photon of 50 MeV.
From the simulation, we find that this EMC threshold discards
effectively the low-energy neutron background
while keeping the photons as many as possible.

The statistic of the $\eta$ meson samples is a key input for
the future experiment and the simulation study in this work.
The statistic depends on the energy setting, the cross section
and the running time of the future experiment.
The energy of proton beam is set at 1.8 GeV, just below the $\rho$ production
threshold to reduce the background.
At this energy, the $\eta$ probability in elastic scattering is around 0.76\%
from the GiBUU simulation of p-$^7$Li collisions. Extrapolated from previous measurements,
the $\eta$ production cross section in p-p collision is about 0.1 mb at 1.8 GeV \cite{Wilkin:2016mfn}.
Thus the cross section in p-A collision is around $0.1\times A$ mb.
The multi-layer target of thin foils (Lithium or Beryllium) will be used in the future experiment
and the luminosity of the fixed target experiment can achieve $10^{35}$ cm$^{-2}$s$^{-1}$.
The light nuclear target is used to reduce the background and the particle multiplicity.
However, considering the event rate capacity of the spectrometer
we assume a conservative event rate of inelastic scattering to be 100 MHz.
To make a conservative estimation of the impact of the future experiment,
we assume the experiment running for just one month with the duty factor of 30\%.
Finally, we estimate that the number of $\eta$ mesons produced in a
future prior experiment is $5.9\times10^{11}$.

In this simulation, we only simulate about 13 millions of inelastic events
of p-A collision, due to the limitations of computing source and storage volume.
In order to predict the sensitivity of the real experiment,
we scale up both the background distributions and
the number of produced $\eta$ samples accordingly,
using a scale-up factor of around $10^5$.
The total event numbers of the future $\eta$ factory experiment will be a spectacular number.

\section{Results and discussions}
\label{sec:result}

From the Monte-Carlo simulations, firstly we estimate the detection efficiencies
of the channels of interests. Secondly, we show the resolutions of the masses
of $\pi^{0}$, $\eta$ and dark scalar particle.
Thirdly, we show the projected background distributions after performing
the event selection criteria.
Fourthly, we compute the upper limits of branching ratios of the studied channels.
Finally, the sensitivities to the model parameters are given from the simulation data.

\subsection{Efficiencies}

\begin{figure}[htbp]
\centering  
\includegraphics[width=0.4\textwidth]{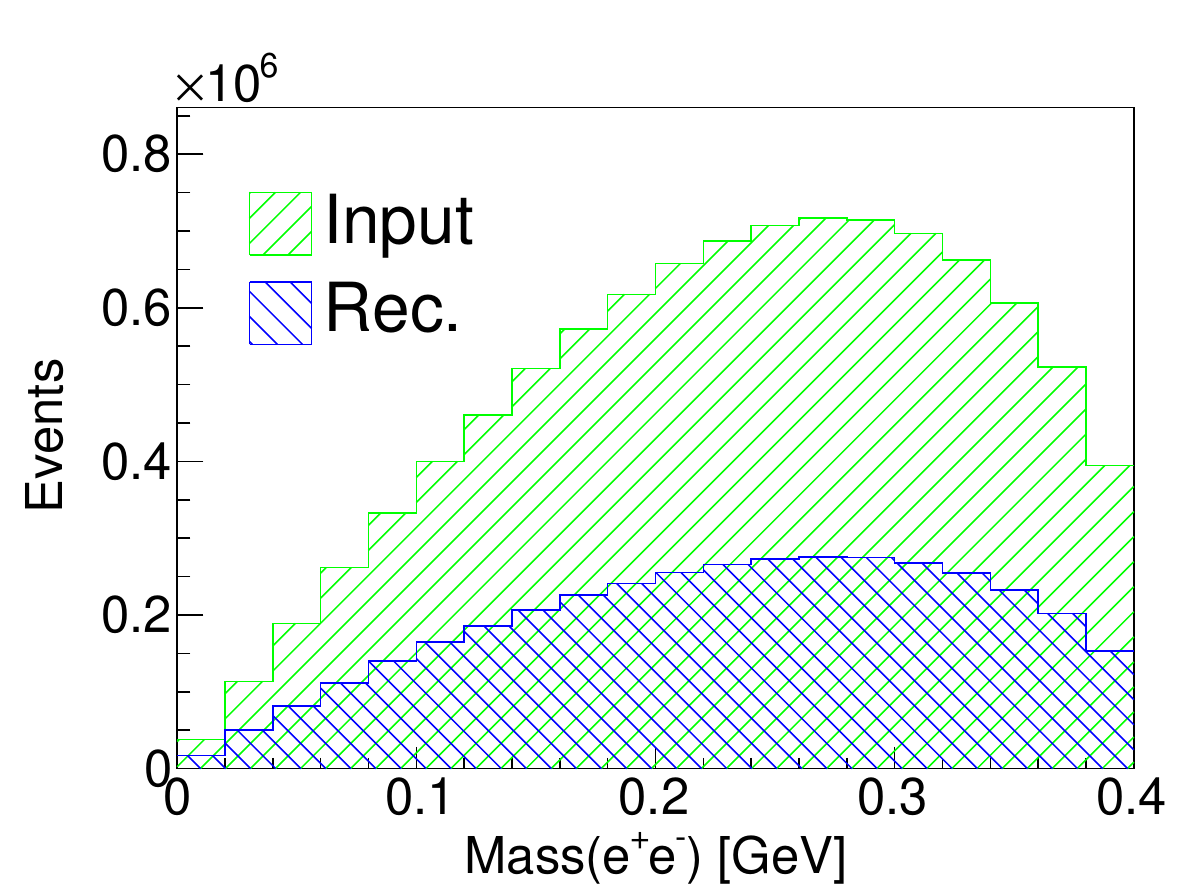}
\caption{The event distributions as a function of the mass of dark scalar particle
for the channel $\eta\rightarrow\pi^0  e^+ e^-$.
The green and blue histograms show respectively the input MC events
from event generator and the reconstructed events from detector simulation. }
\label{fig:MC_and_REC_distributions_EEPi0_channel}
\end{figure}

\begin{figure}[htbp]
\centering  
\includegraphics[width=0.4\textwidth]{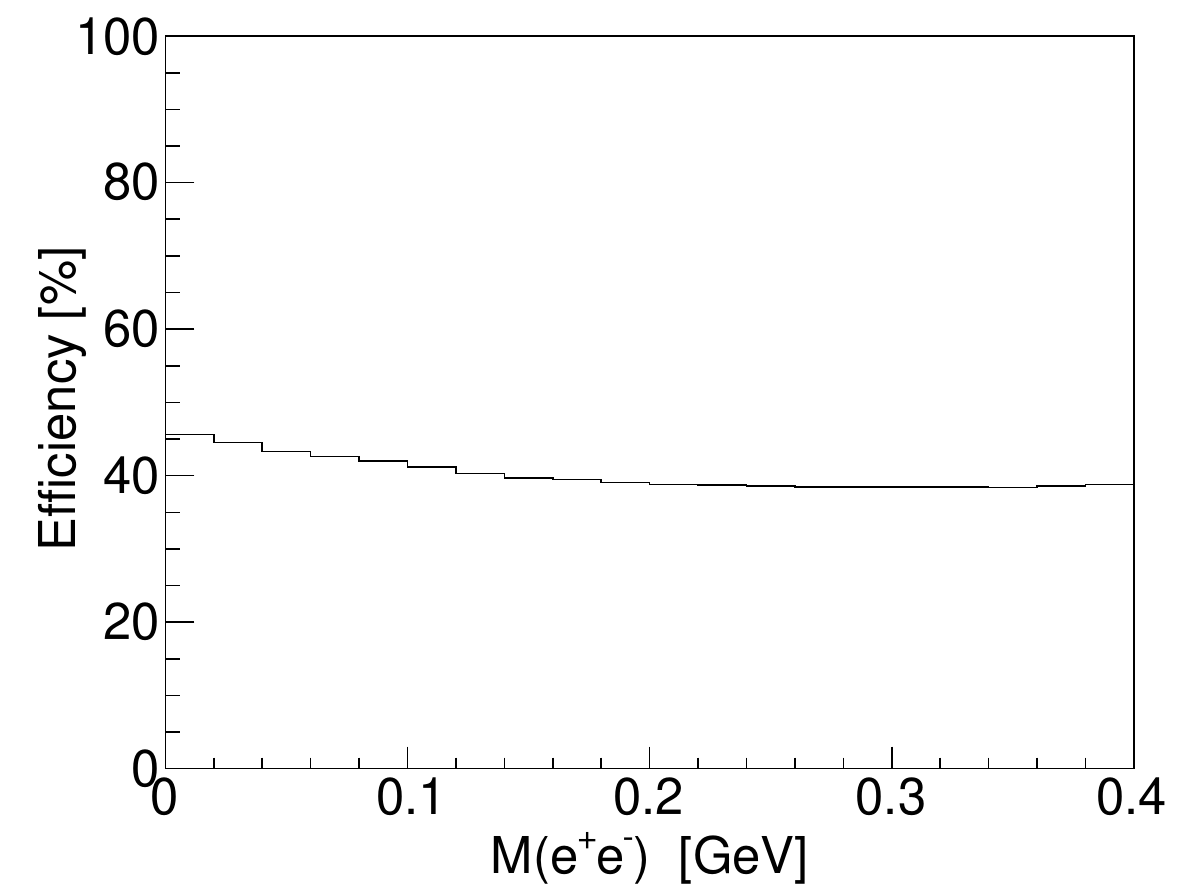}
\caption{The collecting efficiency of the channel $\eta\rightarrow\pi^0  e^+ e-$
as a function of $M(e^+e^-)$.  }
\label{fig:efficiency_EEPi0_channel}
\end{figure}

\begin{figure}[htbp]
\centering  
\includegraphics[width=0.4\textwidth]{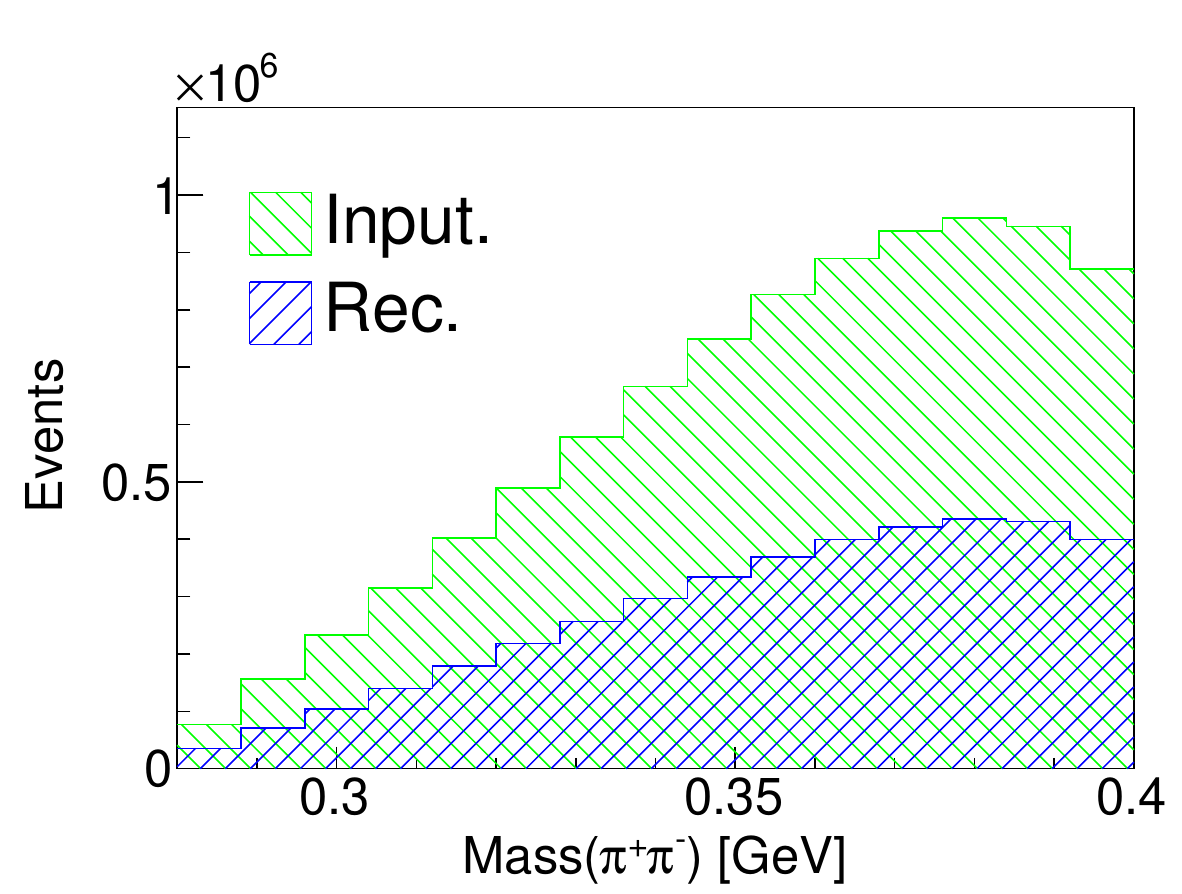}
\caption{The event distributions as a function of the mass of dark scalar particle
for the channel $\eta\rightarrow\pi^0 \pi^+ \pi^-$.
The green and blue histograms show respectively the input MC events
from event generator and the reconstructed events from detector simulation.  }
\label{fig:MC_and_REC_distributions_3pi_channel}
\end{figure}

\begin{figure}[htbp]
\centering  
\includegraphics[width=0.4\textwidth]{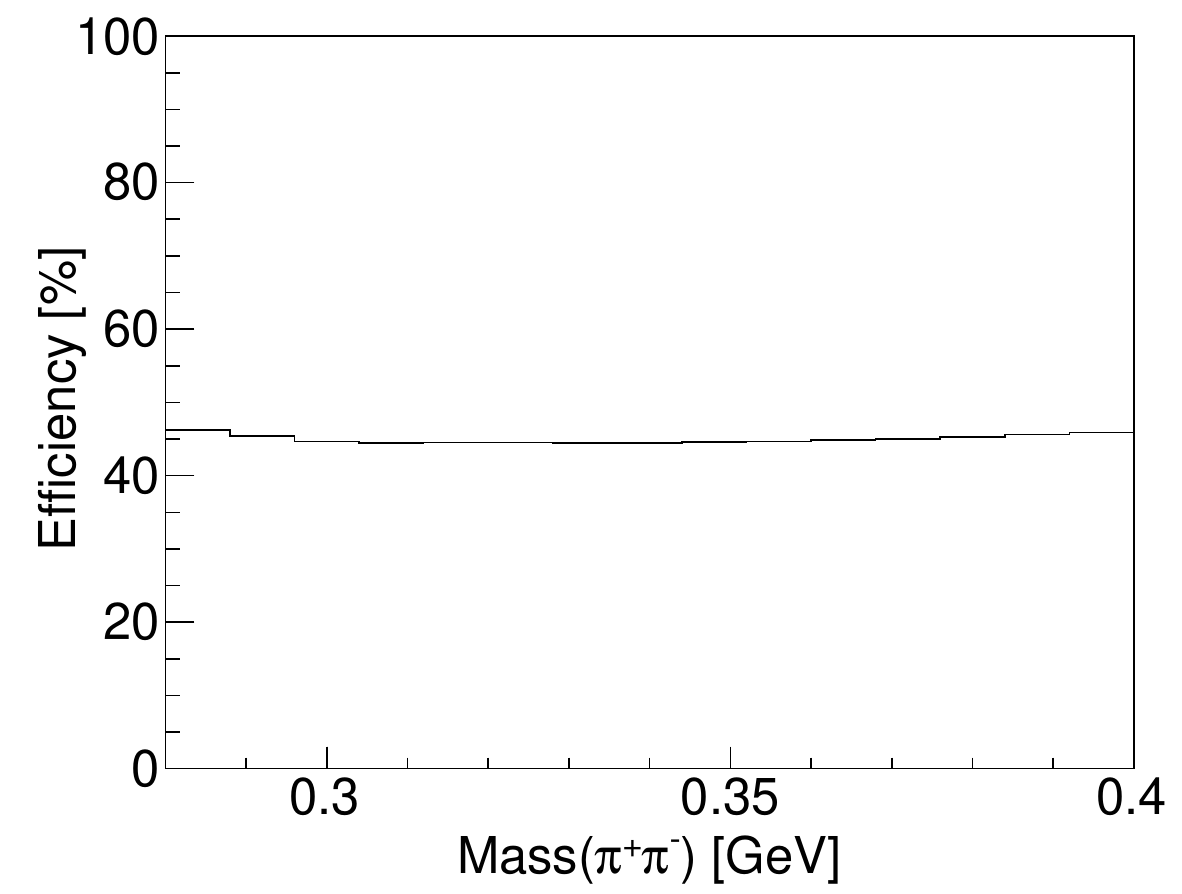}
\caption{The collecting efficiency of the channel $\eta\rightarrow\pi^0 \pi^+ \pi^-$
as a function of $M(\pi^+\pi^-)$.  }
\label{fig:efficiency_3pi_channel}
\end{figure}

The detection efficiencies for the aimed $\eta$ decay channels are the important
quantities in optimizing the design of the spectrometer.
The input MC events and the reconstructed events are shown
in Figs. \ref{fig:MC_and_REC_distributions_EEPi0_channel}
and \ref{fig:MC_and_REC_distributions_3pi_channel},
for the channels $\eta\rightarrow\pi^0  e^+ e^-$ and $\eta\rightarrow\pi^0 \pi^+ \pi^-$ respectively.
One sees that the efficiencies are above 40\% for both channels aimed
for the dark scalar particle exploration.
These efficiencies are satisfying, as they are very close to the pure geometrical acceptances.

\subsection{Invariant mass resolutions}

\begin{figure}[htbp]
\centering  
\includegraphics[width=0.4\textwidth]{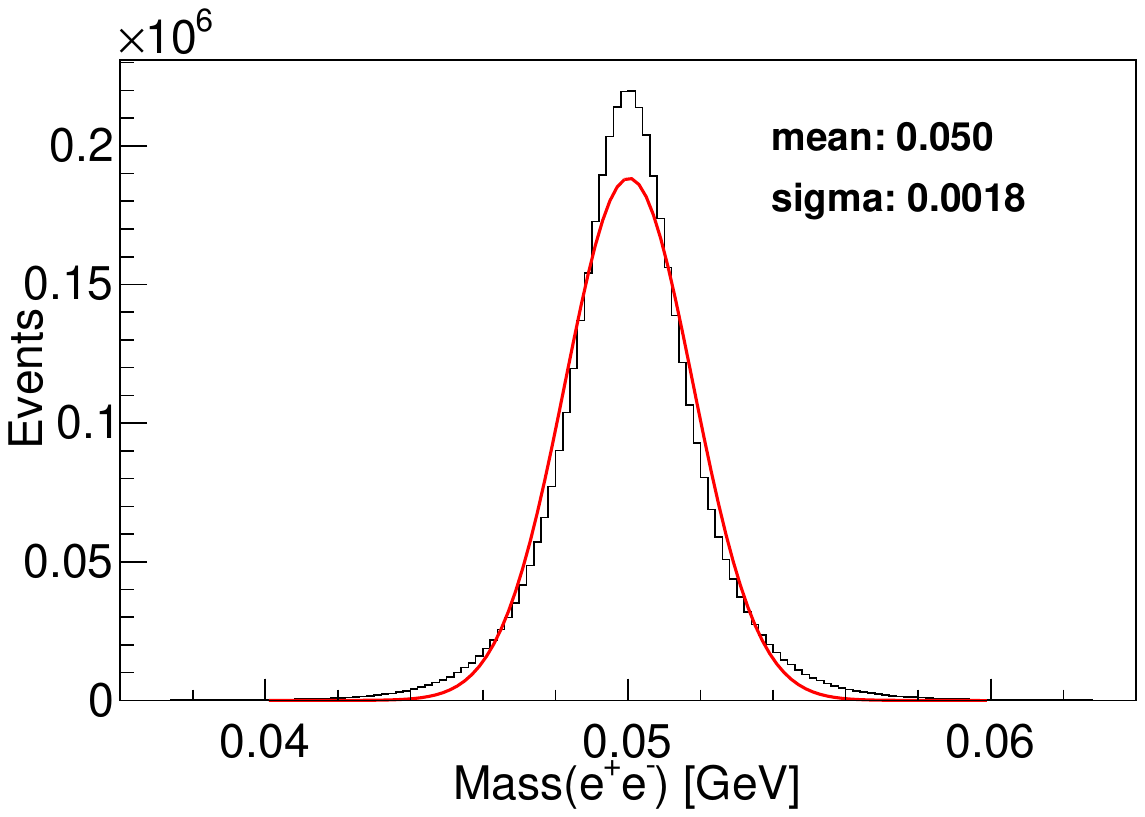}
\caption{The invariant mass distribution of $e^+e^-$ from the dark scalar decay channel
$\eta\rightarrow S\pi^0 \rightarrow e^+ e^-\gamma\gamma$.
The mass of dark scalar particle is assumed to be 50 MeV in the simulation.   }
\label{fig:dark_scalar_mass_distri_EEPi0_channel}
\end{figure}

\begin{figure}[htbp]
\centering  
\includegraphics[width=0.4\textwidth]{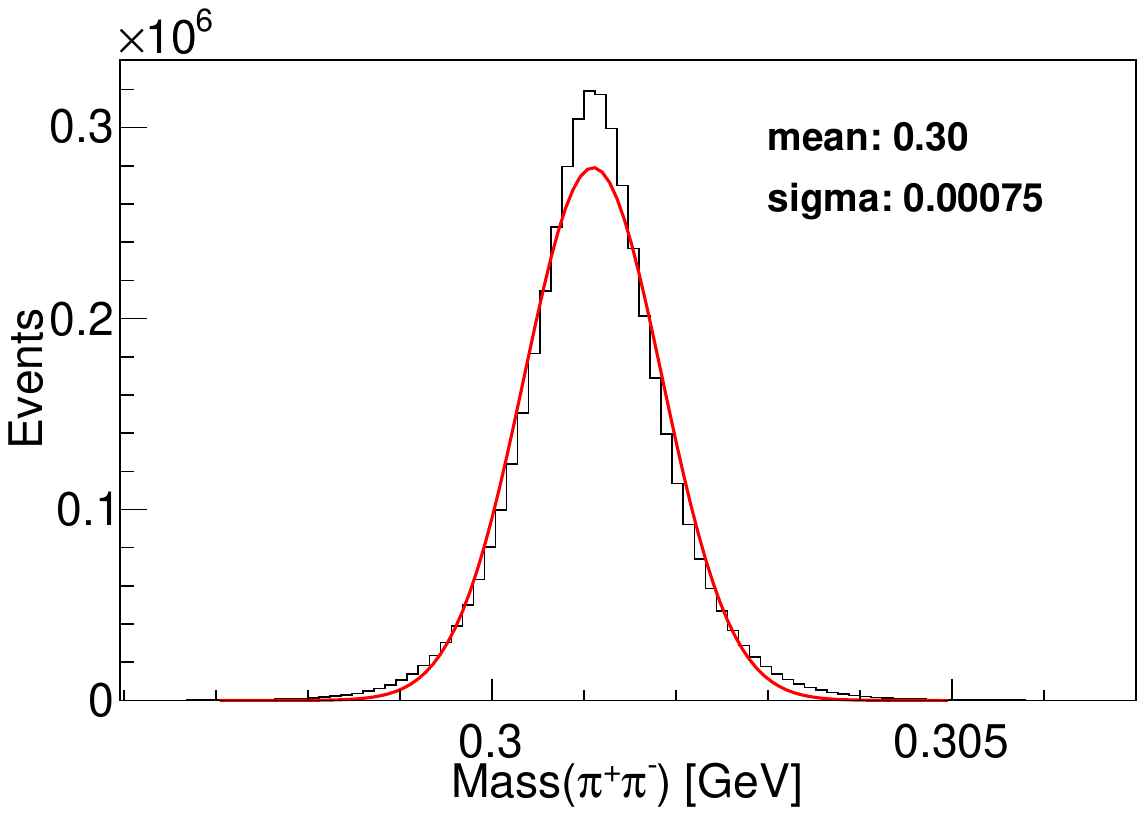}
\caption{The invariant mass distribution of $\pi^+\pi^-$ from the dark scalar decay channel
$\eta\rightarrow S\pi^0 \rightarrow \pi^+\pi^-\gamma\gamma$.
The mass of dark scalar particle is assumed to be 300 MeV in the simulation.   }
\label{fig:dark_scalar_mass_distri_3pi_channel}
\end{figure}

\begin{figure}[htbp]
\centering  
\includegraphics[width=0.4\textwidth]{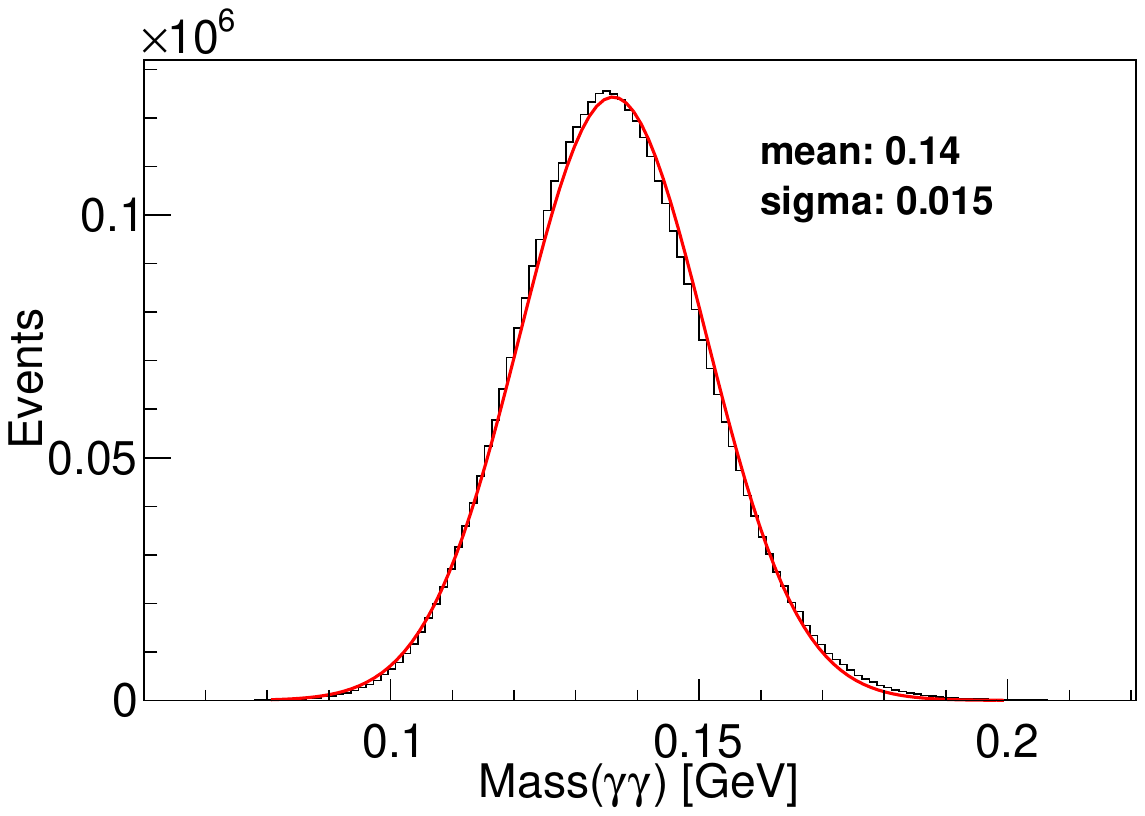}
\caption{The distribution of the reconstructed $\pi^0$ mass
from the channel $\eta\rightarrow e^+ e^-\pi^0 \rightarrow e^+ e^-\gamma\gamma$.   }
\label{fig:pi0_mass_distri_EEPi0_channel}
\end{figure}

\begin{figure}[htbp]
\centering  
\includegraphics[width=0.4\textwidth]{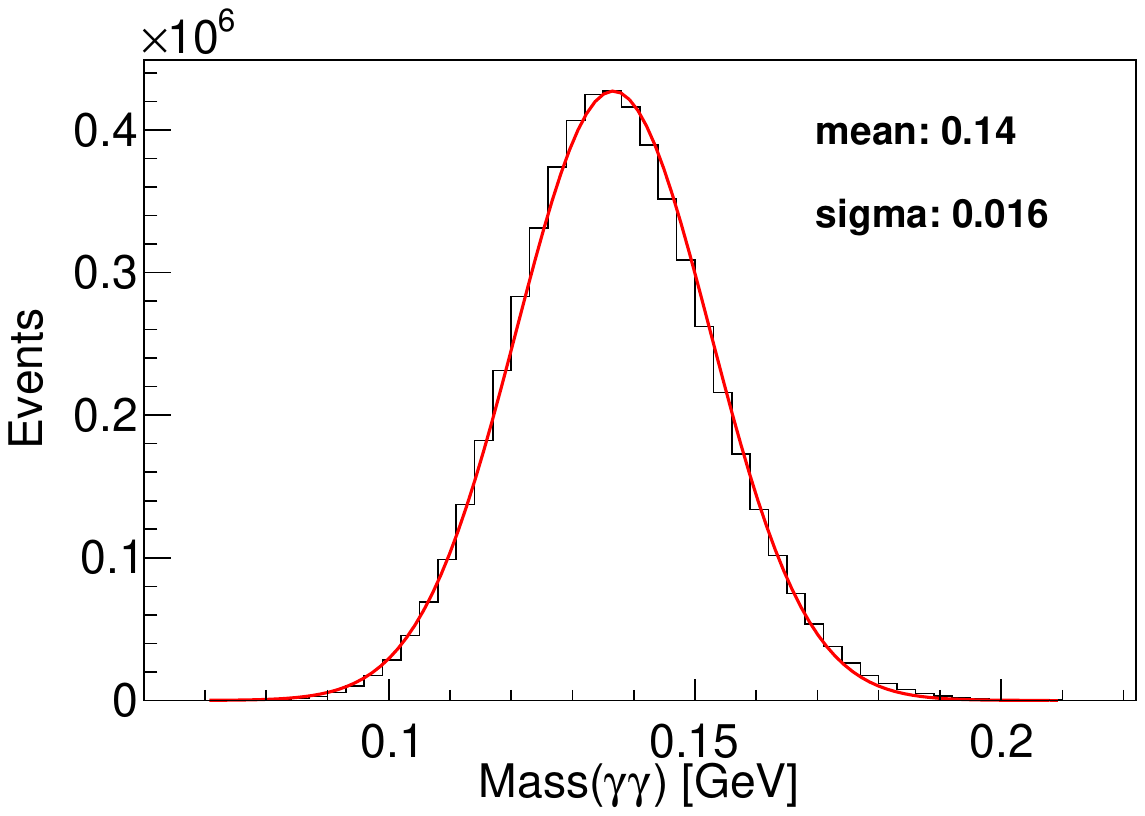}
\caption{The distribution of the reconstructed $\pi^0$ mass
from the channel $\eta\rightarrow \pi^+ \pi^-\pi^0 \rightarrow \pi^+ \pi^-\gamma\gamma$.   }
\label{fig:pi0_mass_distri_3pi_channel}
\end{figure}

\begin{figure}[htbp]
\centering  
\includegraphics[width=0.4\textwidth]{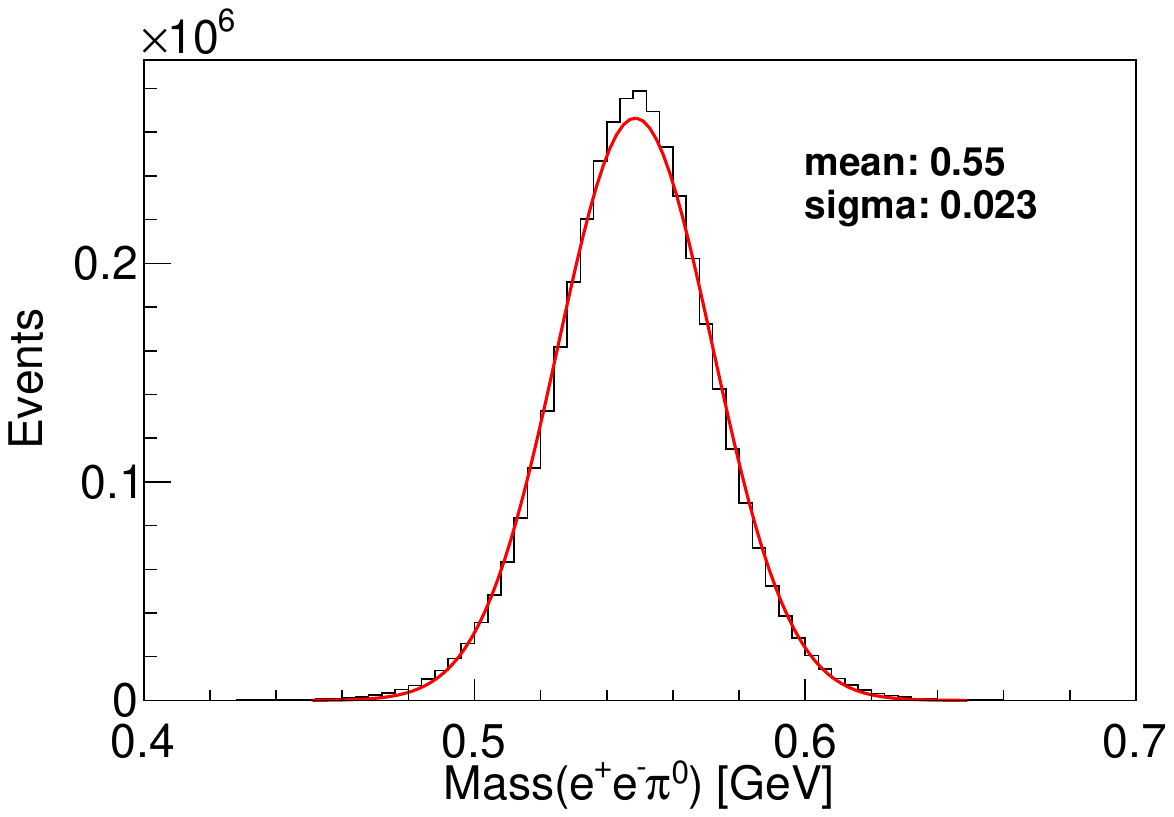}
\caption{The distribution of the reconstructed $\eta$ mass
from the channel $\eta\rightarrow e^+ e^-\pi^0$.  }
\label{fig:eta_mass_distri_EEPi0_channel}
\end{figure}

\begin{figure}[htbp]
\centering  
\includegraphics[width=0.4\textwidth]{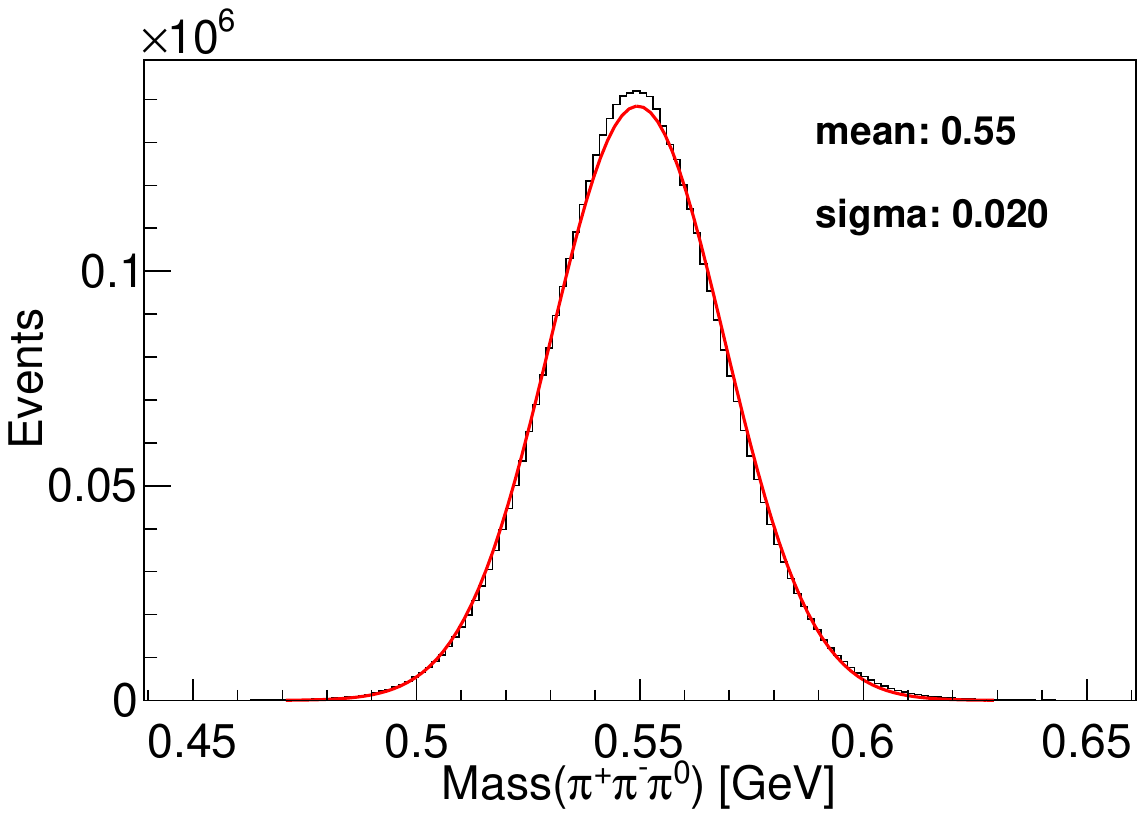}
\caption{The distribution of the reconstructed $\eta$ mass
from the channel $\eta\rightarrow \pi^+ \pi^-\pi^0$.  }
\label{fig:eta_mass_distri_3pi_channel}
\end{figure}

In the simulation, we programmed the decay chains of $\eta$ meson decay
with a presumed dark scalar particle, $\eta\rightarrow S\pi^0 \rightarrow e^+ e^-\gamma\gamma$
and $\eta\rightarrow S\pi^0 \rightarrow \pi^+\pi^-\gamma\gamma$.
From the ChnsRoot simulations, the distributions of the reconstructed mass
of dark scalar particle are shown in Figs. \ref{fig:dark_scalar_mass_distri_EEPi0_channel}
and \ref{fig:dark_scalar_mass_distri_3pi_channel} for $e^+e^-$ and $\pi^+\pi^-$ channels respectively.
Thanks to the small spatial resolution of silicon pixel detector,
the mass resolution for the dark scalar particle is quite small,
less than 2 MeV for both channels.
The small mass resolution is important for the sensitivity of new particle,
as the background events under a narrower peak are fewer.

For the event selection, we also need to identify $\pi^0$ and $\eta$ particles
from the invariant mass distributions.
Based on the ChnsRoot simulations, the distributions of the reconstructed mass
of the decay $\pi^0$ are shown in Figs. \ref{fig:pi0_mass_distri_EEPi0_channel}
and \ref{fig:pi0_mass_distri_3pi_channel}
for $e^+e^-\pi^0$ and $\pi^+\pi^-\pi^0$ channels respectively.
According to the current design of EMC, the mass resolution of $\pi^0$ is not that good,
around 15 MeV for both $\eta$ decay channels.
The distributions of the reconstructed mass
of $\eta$ meson are shown in Figs. \ref{fig:eta_mass_distri_EEPi0_channel}
and \ref{fig:eta_mass_distri_3pi_channel}
for $e^+e^-\pi^0$ and $\pi^+\pi^-\pi^0$ channels respectively.
From our simulation, the mass resolution of $\eta$ meson is around 20 MeV for both studied decay channels.
The resolution of $\eta$ mass mainly comes from the resolution of $\pi^0$,
for our designed spectrometer excels in measuring precisely the momentum of charged particle.
Small mass resolution allows us applying strict criteria for $\pi^0$ and $\eta$ selections,
so as to reduce the background and improve the sensitivity to dark scalar particle.
Improving the energy resolution of EMC is one effective way
to improve the resolutions of the masses of $\pi^0$ and $\eta$.

\subsection{Background distributions}

\begin{figure}[htbp]
\centering  
\includegraphics[width=0.4\textwidth]{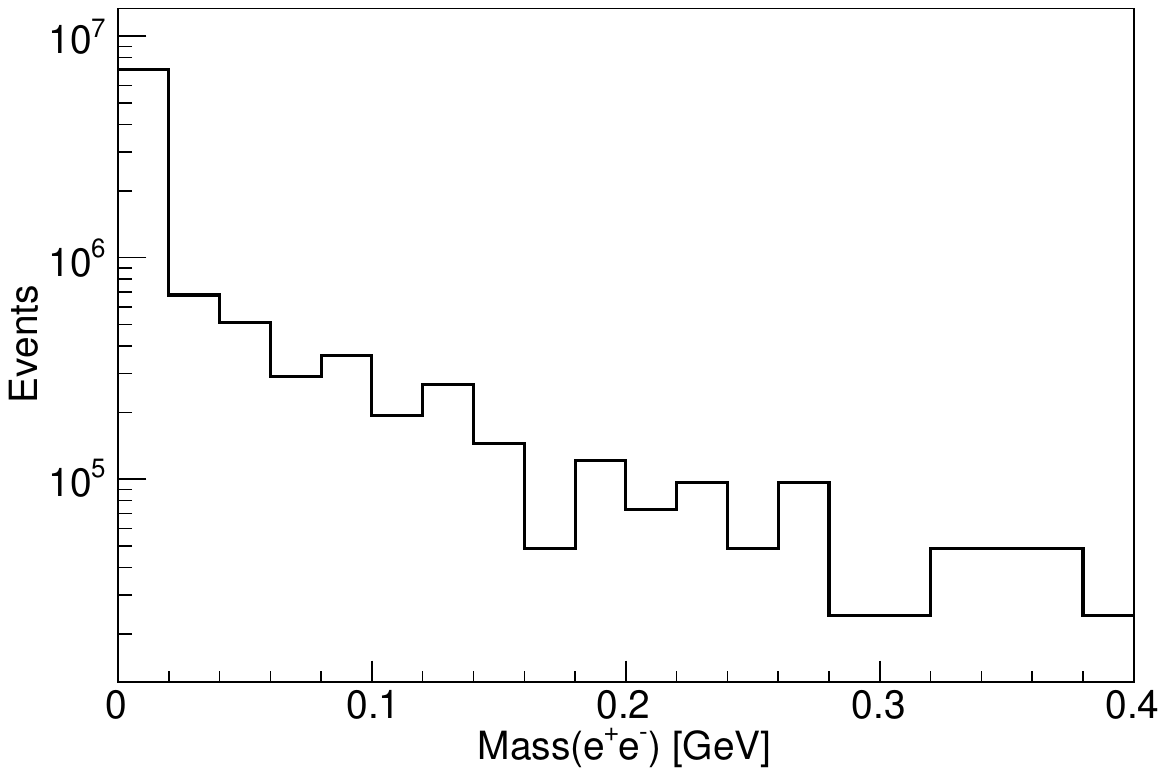}
\caption{The projected invariant mass distribution of $e^+e^-$
in the channel $\eta\rightarrow e^+e^-\pi^0$,
for the suggested one-month running experiment.  }
\label{fig:EpEm_mass_distribution_EEPi0_channel}
\end{figure}

\begin{figure}[htbp]
\centering  
\includegraphics[width=0.4\textwidth]{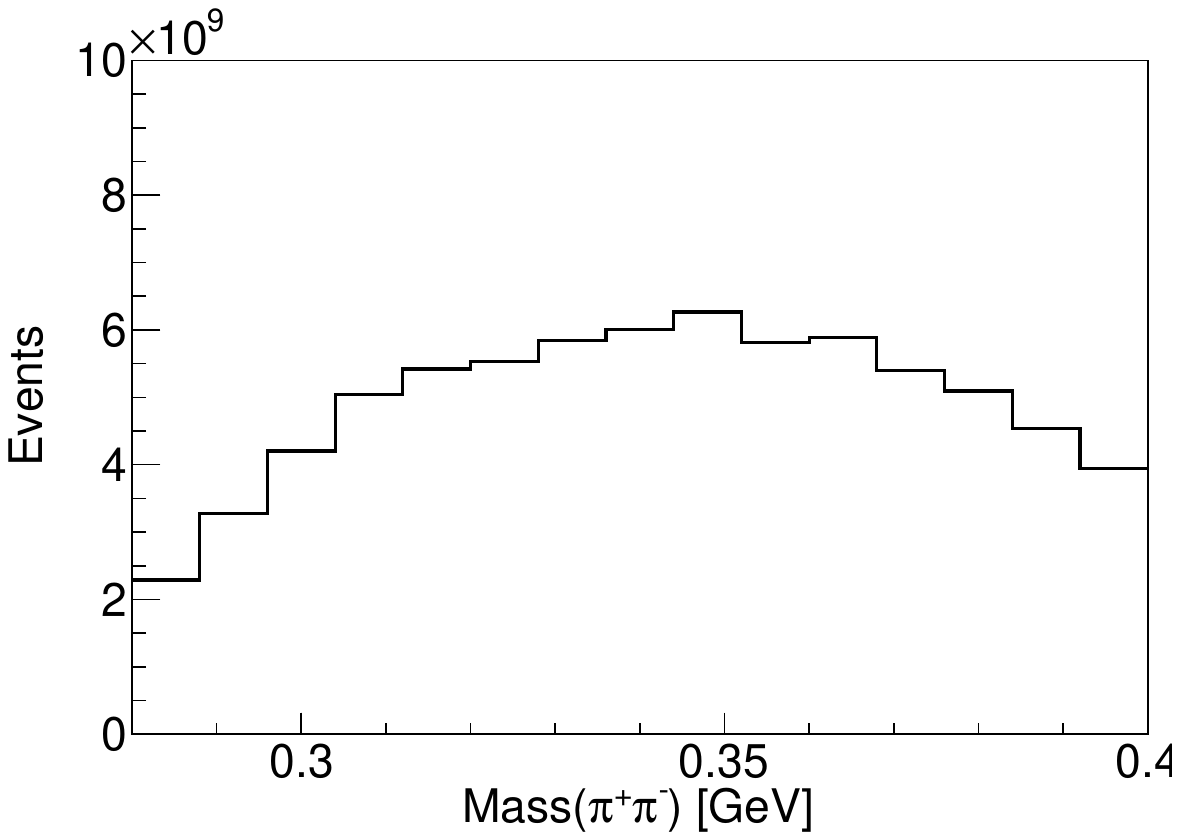}
\caption{The projected invariant mass distribution of $\pi^+\pi^-$
in the channel $\eta\rightarrow \pi^+\pi^-\pi^0$,
for the suggested one-month running experiment.  }
\label{fig:PipPim_mass_distribution_3pi_channel}
\end{figure}

The targeted decay channels of $\eta$ in searching the dark scalar particle are
$\eta\rightarrow e^+e^-\pi^0$ and $\eta\rightarrow \pi^+\pi^-\pi^0$.
The technique to find the dark scalar particle is to look for
a bump in the invariant mass distributions of $e^+e^-$ and $\pi^+\pi^-$.
Before generating the aimed invariant mass distributions,
we need to select the channels of interests.
The reconstructed masses of $\eta$ and $\pi^0$ are required
to be within $\pm 3\sigma$ range.

Figs. \ref{fig:EpEm_mass_distribution_EEPi0_channel} and
\ref{fig:PipPim_mass_distribution_3pi_channel} show the simulated invariant mass distributions
of $e^+e^-$ and $\pi^+\pi^-$, from the channels $\eta\rightarrow e^+e^-\pi^0$
and $\eta\rightarrow \pi^+\pi^-\pi^0$ respectively.
The bin width of histogram is taken as six times of the resolution of dark scalar particle,
so that the dark scalar particle mostly shows up in only one bin.
To make a conservative estimation, in the detector simulation,
the neutrons above the hit threshold of EMC are all misidentified as photons.
In the invariant mass distributions of $e^+e^-$ and $\pi^+\pi^-$,
one sees no peaks, because there is no dark scalar particle implemented
in the GiBUU event generator.
Therefore, the obtained invariant mass distributions depict
only the background distributions without the dark scalar particle.
The lower the background distribution, the better sensitivity of the experiment will be.

\subsection{Branching-ratio upper limits}

\begin{figure}[htbp]
\centering  
\includegraphics[width=0.4\textwidth]{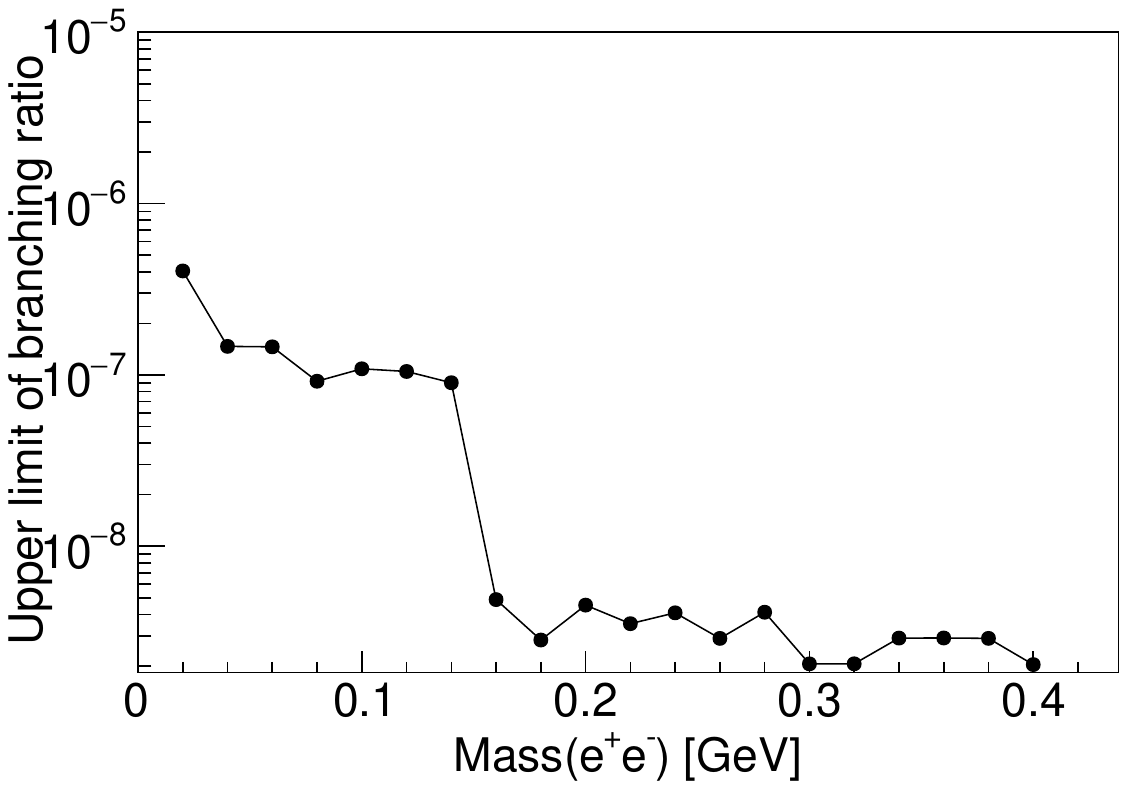}
\caption{The projected branching-ratio upper limit of dark scalar particle
in the decay channel $\eta\rightarrow S\pi^0 \rightarrow e^+ e^-\gamma\gamma$,
for the suggested experiment of one-month running.  }
\label{fig:Br_upper_limit_EEPi0_channel}
\end{figure}

\begin{figure}[htbp]
\centering  
\includegraphics[width=0.4\textwidth]{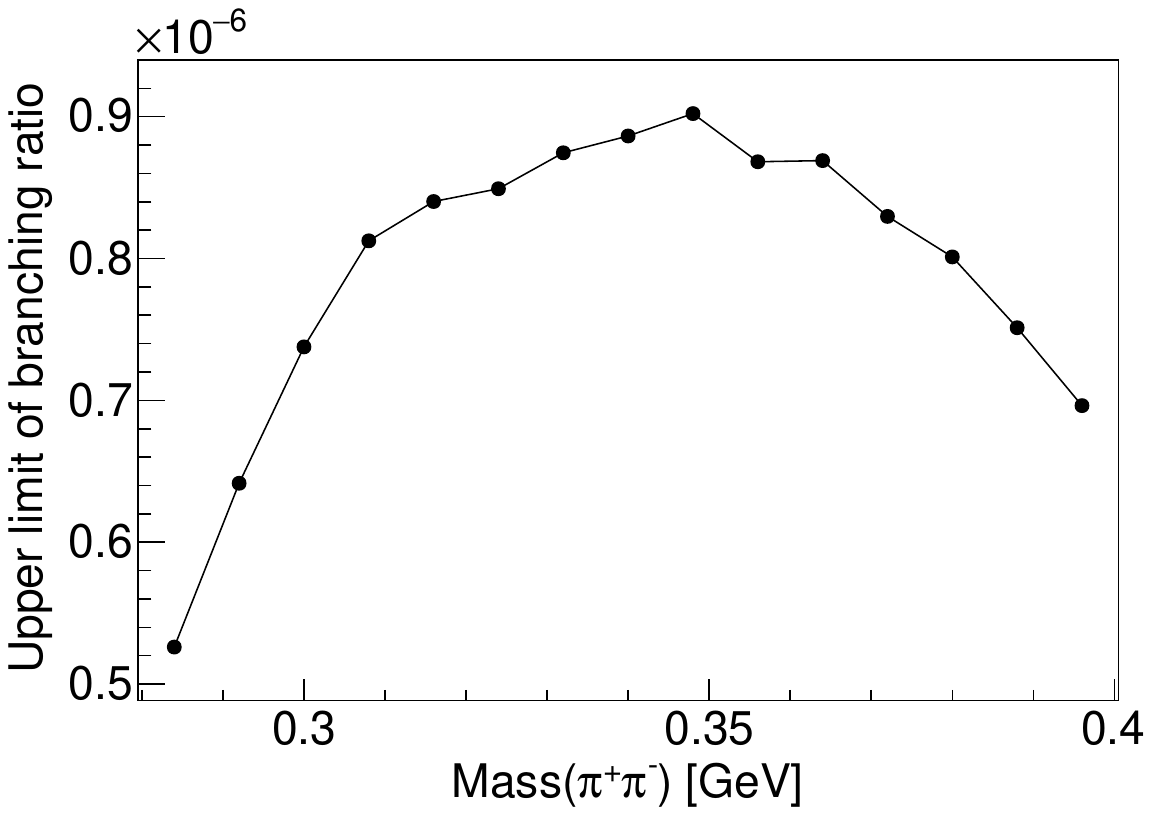}
\caption{The projected branching-ratio upper limit of dark scalar particle
in the decay channel $\eta\rightarrow S\pi^0 \rightarrow \pi^+\pi^-\gamma\gamma$,
for the suggested experiment of one-month running.   }
\label{fig:Br_upper_limit_3pi_channel}
\end{figure}

The invariant mass distributions of $e^+e^-$ and $\pi^+\pi^-$,
present no bump, as they are just the background distributions
without the dark scalar particle in the decay.
Since there is no signal peak in the distribution,
the significance of dark scalar particle is less than $3\sigma$.
With the background distribution after the event selection process,
we can estimate the branching-ratio upper limit of the dark scalar decay channel.
The branching-ratio upper limit of a new particle in the decay is simply given by,
\begin{equation}
\begin{split}
{\rm Br.~upper~limit} = \frac{3\times\sqrt{N_{\rm bg}^{i}}}{N_{\eta}\times\epsilon_{i}},
\end{split}
\label{eq:BrUpperLimit}
\end{equation}
where $N_{\rm bg}^{i}$ is the resulting number of background events in bin i,
$N_{\eta}$ is total number of $\eta$ mesons produced in the experiment,
and $\epsilon_{i}$ is the efficiency of detecting the dark scalar particle in the mass bin i.
The confidence level is at 99 \% for the upper limit estimated with Eq. (\ref{eq:BrUpperLimit}).
The statistic of total $\eta$ samples is discussed at the beginning of this section.

Figs. \ref{fig:Br_upper_limit_EEPi0_channel} and \ref{fig:Br_upper_limit_3pi_channel}
show the branching-ratio upper limits of a dark scalar particle in the decay
as a function of the particle mass, in the channels $\eta\rightarrow S\pi^0 \rightarrow e^+ e^-\gamma\gamma$
and $\eta\rightarrow S\pi^0 \rightarrow \pi^+\pi^-\gamma\gamma$ respectively.
In Fig. \ref{fig:Br_upper_limit_EEPi0_channel}, one sees a fast drop of the upper limit
around 0.14 GeV. This is because most of the background electrons
in the simulation are from the decay of $\pi^0$.
In the large mass region above the pion mass,
the projected upper limit of dark scalar particle is close to $10^{-9}$
in the $e^+e^-$ channel.
In the $\pi^+\pi^-$ channel, the branching-ratio upper limit
of dark scalar particle is below $10^{-6}$.
As the direct $\pi^+\pi^-\pi^0$ decay is one of the main decay channels of $\eta$,
the upper limit of dark scalar particle given in this channel is not that small.

\subsection{Sensitivities to model parameters}

\begin{figure}[htbp]
\centering  
\includegraphics[width=0.4\textwidth]{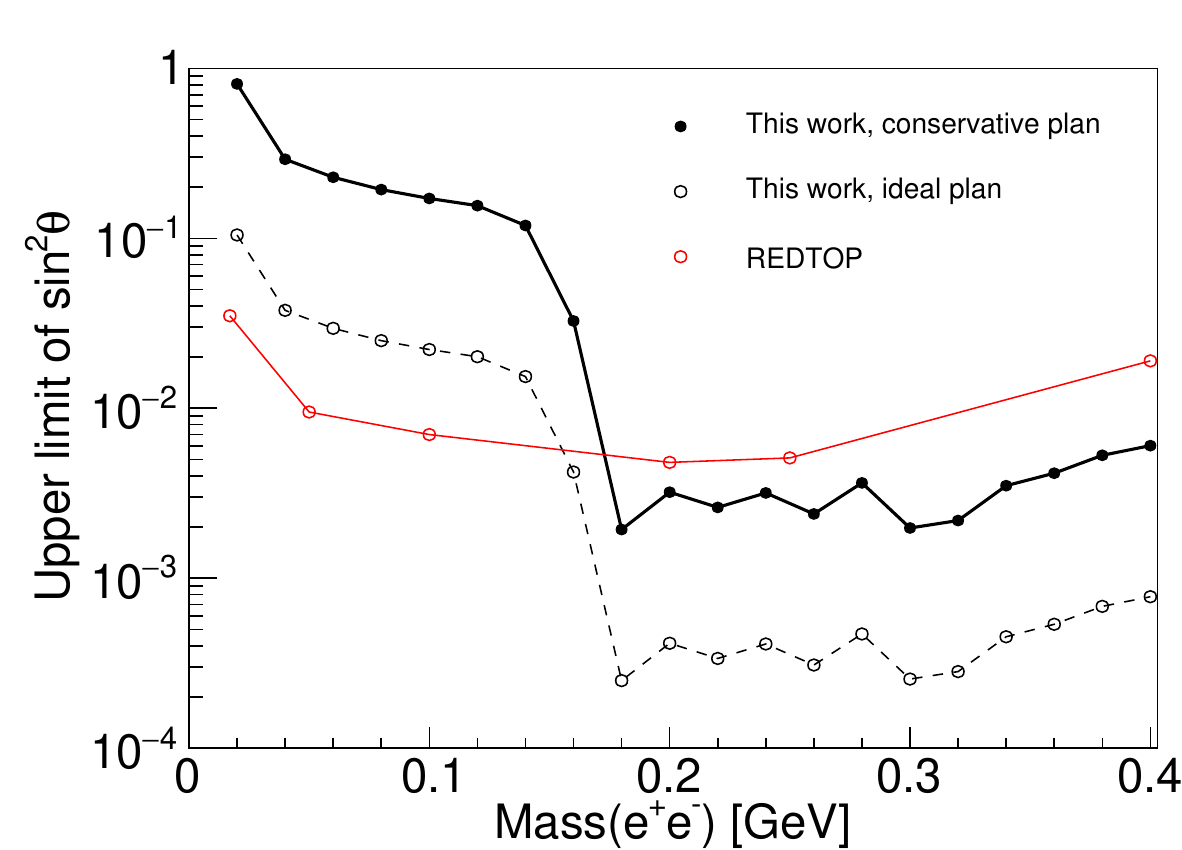}
\caption{The sensitivity to the parameter $\sin^{2}\theta$ in the minimal scalar model as a function of
the mass of dark scalar particle, for the suggested experiment of one-month running (black solid curve,
the conservative plan).
The projected sensitivity for an ideal case is also shown in the figure (black dashed curve).
The ideal experimental plan is for one-year running at event rate of 500 MHz.
The red dashed curve shows the preliminary result of REDTOP experiment \cite{REDTOP:2022slw}.  }
\label{fig:theta2_sensitivity}
\end{figure}

\begin{figure}[htbp]
\centering  
\includegraphics[width=0.4\textwidth]{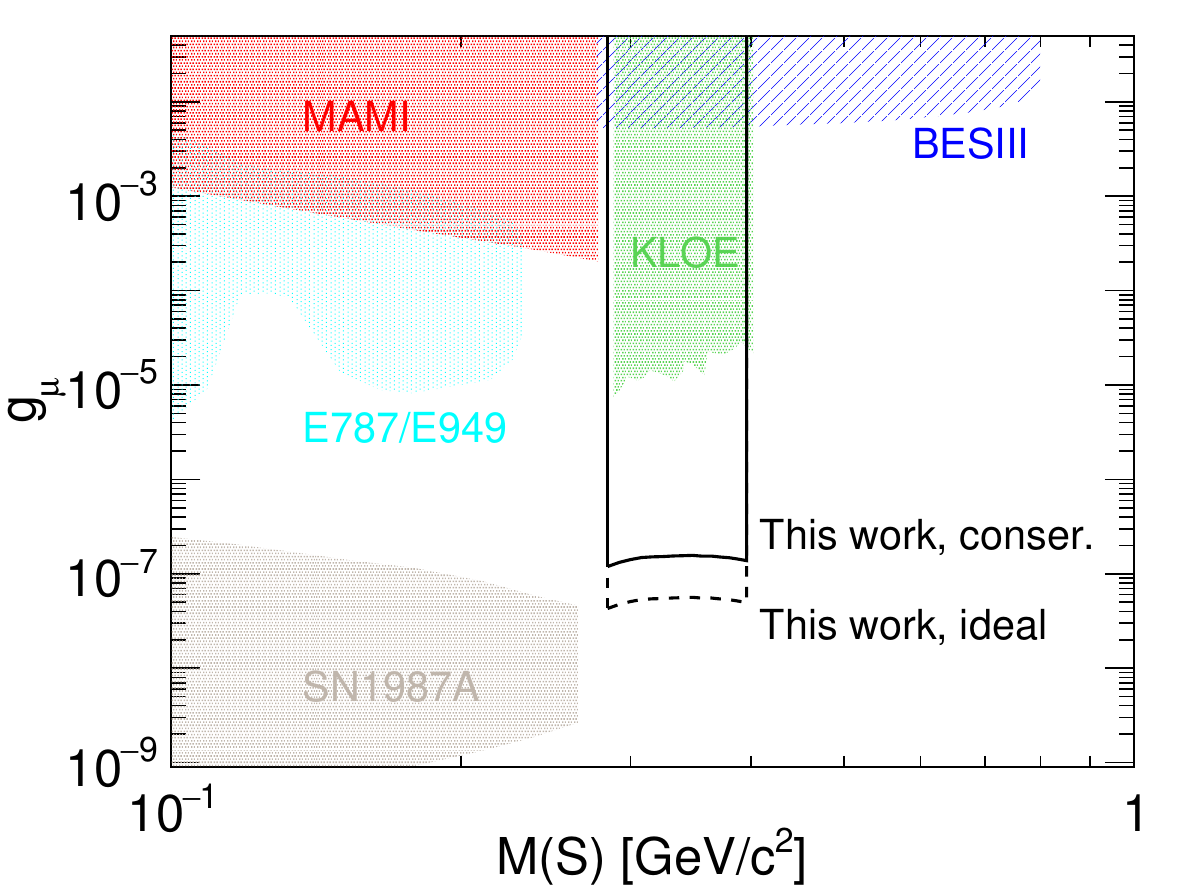}
\caption{The sensitivity to the parameter $g_{u}$ in the hadrophilic scalar model as a function of
the mass of dark scalar particle, for the suggested experiment of one-month running
(black solid curve, the conservative plan).
The projected sensitivity for an ideal case is also shown in the figure (black dashed curve).
The ideal experimental plan is for one-year running at event rate of 500 MHz.
Some previous experimental data for the constraints are taken from
E787/E949 \cite{E787:2002qfb,E787:2004ovg,E949:2007xyy,BNL-E949:2009dza},
MAMI \cite{Liu:2018qgl}, BESIII \cite{BESIII:2016tdb},
KLOE \cite{KLOE-2:2016zfv}, SN 1987A \cite{Batell:2018fqo}.    }
\label{fig:gu_sensitivity}
\end{figure}

Applying the model description of the dark scalar particle in $\eta$ rare decay,
the branching-ratio upper limit of $\eta$ to dark scalar particle
can be used to constrain the free parameters in the model.
The sensitivity to the model parameter is the precision of the parameter
on which level we can test the model with satisfying significance.
In an experiment, the sensitivity to the model parameter is closely
related to the measured upper limit of the branching ratio,
via Eq. (\ref{eq:BrAndTheta}) or Eq. (\ref{eq:BrAndgu}).

Fig. \ref{fig:theta2_sensitivity} shows the projected sensitivity of
the mixing angle parameter as a function of the mass of dark scalar particle,
under the minimal scalar model.
The sensitivity is based on the projected upper limit of branching ratio
of $\eta\rightarrow S\pi^0 \rightarrow e^+ e^-\gamma\gamma$,
for a prior experiment of one-month running.
One sees that the sensitivity to $\theta$ is around the level of $10^{-1}$
at the confidence level of 99\%.
REDTOP's preliminary projection is also shown for comparisons \cite{REDTOP:2022slw}.
In the small-mass region our result is worse than REDTOP's result,
while in the large-mass region our result is as similar as REDTOP's result.
In Fig. \ref{fig:theta2_sensitivity}, we also present the sensitivity projection
for an ideal experimental plan of one-year running at event rate of 500 MHz.

Fig. \ref{fig:gu_sensitivity} shows the projected sensitivity of
the coupling parameter $g_u$ as a function of the mass of dark scalar particle,
under the hadrophilic scalar model.
The sensitivity is based on the projected upper limit of branching ratio
of $\eta\rightarrow S\pi^0 \rightarrow \pi^+\pi^-\gamma\gamma$,
for a prior experiment of one-month running.
One sees that the sensitivity to $g_u$ is close to $10^{-6}$
at the confidence level of 99\%.
One finds that even with only one-month running of our suggested experiment,
the sensitivity will surpass the existing experiments in the corresponding mass domain.
The current experimental constraints by MAMI and BESIII are
from the analyses of $\eta\rightarrow \pi^0 \gamma\gamma$
and $\eta^{\prime}\rightarrow \pi^0\pi^+\pi^-$ data, respectively,
under the same hadrophilic scalar model.
In Fig. \ref{fig:gu_sensitivity}, we also present the sensitivity projection
for an ideal experimental plan of one-year running at event rate of 500 MHz.
With years running of Huizhou $\eta$ factory program,
our constraints on the hardrophilic scalar model
will be similar to the proposed REDTOP experiment \cite{REDTOP:2022slw}.

\section{Summary}
\label{sec:summary}

A super $\eta$ meson factory at Huizhou is proposed to
explore the new physics and to precisely test the SM.
The total number of $\eta$ events for a prior experiment of one-month running
is estimated to be at the level of $10^{11}$,
while the total number of inelastic scattering events is
estimated to be at the level of $10^{13}$.
The cross section of $\eta$ meson production
in $p$-$A$ collision is given by the GiBUU event generator.

To study the performance of the conceptual design of the spectrometer
and investigate the physics impacts of the suggested experiments,
we have constructed a simulation framework for the experiment.
Both the signal processes ($\eta\rightarrow S\pi^0 \rightarrow e^+ e^-\gamma\gamma$
and $\eta\rightarrow S\pi^0 \rightarrow \pi^+\pi^-\gamma\gamma$)
and the background processes ($\eta\rightarrow e^+ e^-\gamma\gamma$
and $\eta\rightarrow \pi^+\pi^-\gamma\gamma$) have been simulated.
The signal events of the dark scalar portal particle in $\eta$ decay
are generated with the simple computer programs coded by us.
The background events are generated by GiBUU event generator.
We also build a detector simulation tool ChnsRoot,
which is based on the FairRoot framework.

From our simulation, we find that the designed spectrometer
has a large efficiency (around 40\%) in collecting the events
of interests ($\eta\rightarrow S\pi^0 \rightarrow e^+ e^-\gamma\gamma$
and $\eta\rightarrow S\pi^0 \rightarrow \pi^+\pi^-\gamma\gamma$).
Thanks to the small spatial resolution of the silicon pixel tracker,
the invariant mass resolution of the dark scalar particle is excellent
($<2$ MeV) and the invariant mass resolution of $\eta$ is acceptable ($\sim 20$ MeV).
The energy resolution of the photon should be improved in the future,
in order to improve further the signal-to-background ratio.
The branching-ratio upper limits of $\eta\rightarrow S\pi^0 \rightarrow e^+ e^-\gamma\gamma$
and $\eta\rightarrow S\pi^0 \rightarrow \pi^+\pi^-\gamma\gamma$ are projected
with our simulation framework.
The branching-ratio upper limit of the dark scalar particle in $e^+e^-$
channel can reach $10^{-9}$ in the larger mass region above the pion mass.
The branching-ratio upper limit of the dark scalar particle in $\pi^+\pi^-$
channel is at the level of $10^{-6}$.
The sensitivities to the parameters of the minimal scalar model
and the hadrophilic model are provided by the simulations as well.

In this simulation study of dark scalar sensitivities,
the experimental uncertainties are not evaluated quantitatively.
Here we provide some basic information for estimating the uncertainties.
The statistical uncertainty will be quite small,
for we are going to collect a huge number of $\eta$ meson samples
at Huizhou $\eta$ factory. The systematic errors are related to
the performances of the detectors, and they are anticipated to
be dominant for the total experimental uncertainty.
The systematic uncertainties for the sensitivity study
are mainly from the beam monitor,
the detection efficiency, the particle misidentification
and the momentum resolutions.
From the experiences of the current high-energy and nuclear experiments,
these systematic uncertainties are under the control
and at the level of several percentages.

\bibliographystyle{apsrev4-1}
\bibliography{refs}

\end{document}